\documentclass[12pt,rotate,psfig,epsf]{aastex}
\usepackage{psfig,rotate}
\title{Timing Studies on RXTE Observations of SAX J2103.5+4545}
\author{A. Baykal$^1$, S.\c{C}. \.{I}nam $^2$, M.J. Stark $^3$, C.M. Heffner $^{3,4}$, A.E. Erkoca $^5$, J.H. Swank $^6$ \\ 
 $^1$ Physics Department,\\ Middle East Technical University, 06531 Ankara,
Turkey \\ 
 altan$@$astroa.physics.metu.edu.tr \\ $^2$ Department of Electrical and
Electronics Engineering,\\ 
 Ba\c{s}kent University, 06530 Ankara, Turkey \\ inam$@$baskent.edu.tr \\
 $^3$ Lafayette College, Easton, PA 18042, USA \\ starkm$@$lafayette.edu \\ $^4$ Department of Physics, University of California \\ Riverside, CA 92521, USA \\ carolyn.heffner$@$email.ucr.edu  
\\ $^5$ Department of Physics, University of Arizona, Tucson, AZ 85721, USA \\
aerkoca$@$physics.arizona.edu \\ $^6$ NASA Goddard Space Flight Center, Greenbelt, MD 20771, USA \\ swank$@$milkyway.gsfc.nasa.gov}

\date{}
\begin{document}

\begin{abstract}

 SAX J2103.5+4545 has been continuously monitored
for $\sim $ 900 days by Rossi X-ray Timing Explorer (RXTE)
 since its outburst in July 2002.
Using these observations and previous
 archival RXTE observations of SAX J2103.5+4545,
we refined the binary orbital parameters and find the new
orbital period as P= (12.66536 $\pm $ 0.00088) days and the
eccentricity as 0.4055$\pm$ 0.0032.
With these new orbital parameters, we constructed the pulse
frequency and pulse frequency derivative histories
of the pulsar and confirmed the correlation between
X-ray flux and pulse frequency derivative presented by
Baykal, Stark and Swank (2002).  We constructed the
power spectra for the fluctuations
of pulse frequency derivatives and found that
the power law index of the noise spectra is 2.13 $\pm$ 0.6. The
power law index is consistent with random walk in pulse
 frequency derivative and is the steepest among the HMXRBs.
 X-ray spectra analysis confirmed the inverse correlation
trend between power-law index and X-ray flux found by
Baykal, Stark and Swank (2002).

{\bf{Keywords:}} X-rays:binaries; Stars:neutron; pulsars:individual:SAX J2103.5+4545; accretion, accretion disks
\end{abstract}
\maketitle
\section{Introduction}

The transient X-ray source SAX J2103.5+4545 was discovered by {\it{BeppoSAX}} 
during its outburst between 1997 February and September with 
358.61s pulsations and a spectrum consistent with an absorbed power law 
model with the photon index of $\sim1.27$ and the absorption column 
density of $\sim3.1\times 10^{22}$cm$^{-2}$ (Hulleman, in't Zand, \& Heise 1998)

Starting with the November 1999 outburst detected by the 
{\it{all-sky monitor (ASM)}} on 
the {\it{Rossi X-ray Timing Explorer (RXTE)}},  SAX J2103.5+4545 has been continuously monitored for more than a year through 
regular pointed {\it{RXTE}} observations. From RXTE observations, 
the orbital period and eccentricity of the orbit were found to be 
12.68 days and 0.4  (Baykal, Stark, \& Swank 2000a,b). 
In the timing analysis, the source was initially 
found to be spinning up for $\sim 150$ days, 
at which point the flux dropped quickly by 
a factor of $\simeq 7$, and a weak spin-down began afterwards 
(Baykal, Stark, \& Swank 2002). Significant  
correlation between X-ray flux and spin-up rate was explained by using the Ghosh \& Lamb (1979) accretion disk model. 
The X-ray spectra well fitted the
absorbed power law model with high energy cutoff and a $\sim $6.4 keV
fluorescent emission line (Baykal et al. 2002). 

Orbital parameters found by using {\it{RXTE}} observations of the source 
(Baykal et al. 2000a,2000b) indicated that the source has a high mass
companion. The optical companion of SAX J2103.5+4545 was recently discovered to be a Be type star with a visual magnitude of $\sim 14.2$  
(Reig, Neguerela, Fabregat et al.2004; Flippova, Lutovinov, Shtykovsky et al.
2004). 

SAX J2103.5+4545 had another outburst on July 2002 and started to be continuously monitored by {\it{RXTE}}. Using $\sim 2$ months of {\it{RXTE}}-PCA observations around 
the simultaneous coverage of {\it{RXTE}} and {\it{XMM-Newton}} on January 6, 
2003, spin period and spin-up rate of the
source were found to be $(354.7940\pm 0.0008)$s  and $(7.4\pm 0.9)\times
10^{-13}$ Hz.s$^{-1}$ (Inam, Baykal, Swank et al. 2004a). Using this simultaneous
coverage, Inam et al. (2004a) discovered 22.7s quasi-periodic oscillations with
an rms fractional amplitude of $\sim6.6$\% and a soft spectral component 
consistent with a blackbody emission with $kT\sim 1.9$ keV and an emission radius of $\sim 0.3$ km.

SAX J2103.5+4545 was also observed with the {\it{INTEGRAL}} observatory in the
3-200 keV band resulting in a significant detection up to $\sim 100$ keV
(Lutovinov, Molkov,\& Revnivtsev 2003; Flippova, Lutoviniv, Shtykovsky et al.
2004; Play, Reig, Martinez Nunez et al. 2004; Falanga, di Salva, Burderi et al. 2005; Sidoli, Mereghetti, Larsson et al. 2005). The spectral parameters found in the {\it{INTEGRAL}} observations of 
the source were found to be compatible with those found by Baykal et al. (2002).

\section{Instrument and Observations}

The dataset in this paper consists of RXTE monitoring observations followed by the November 1999 and July 2002 outbursts of SAX J2103.5+4545. The results presented here are based on data collected with the Proportional 
Counter Array (PCA, Jahoda et al., 1996). The PCA instrument consists of an array of five collimated xenon/methane multianode proportional counters. The total effective
 area is approximately 6250 cm$^{2}$ and the field of view is $\sim 1^{0}$ FWHM. The nominal energy range extends from 2 to 60 keV. The number of active PCUs varied between 1 and 5 during the observations. Observations after 2000 May 13 belong to the observational epoch for which background level for one of the PCUs (PCU0) increased due to the fact that this PCU started to operate without a propane layer. Latest combined background models (CMs) were used together with FTOOLS 6.0 to estimate the appropriate background for timing and spectral analysis.

\begin{figure}[tb]
\begin{center}
\psfig{file=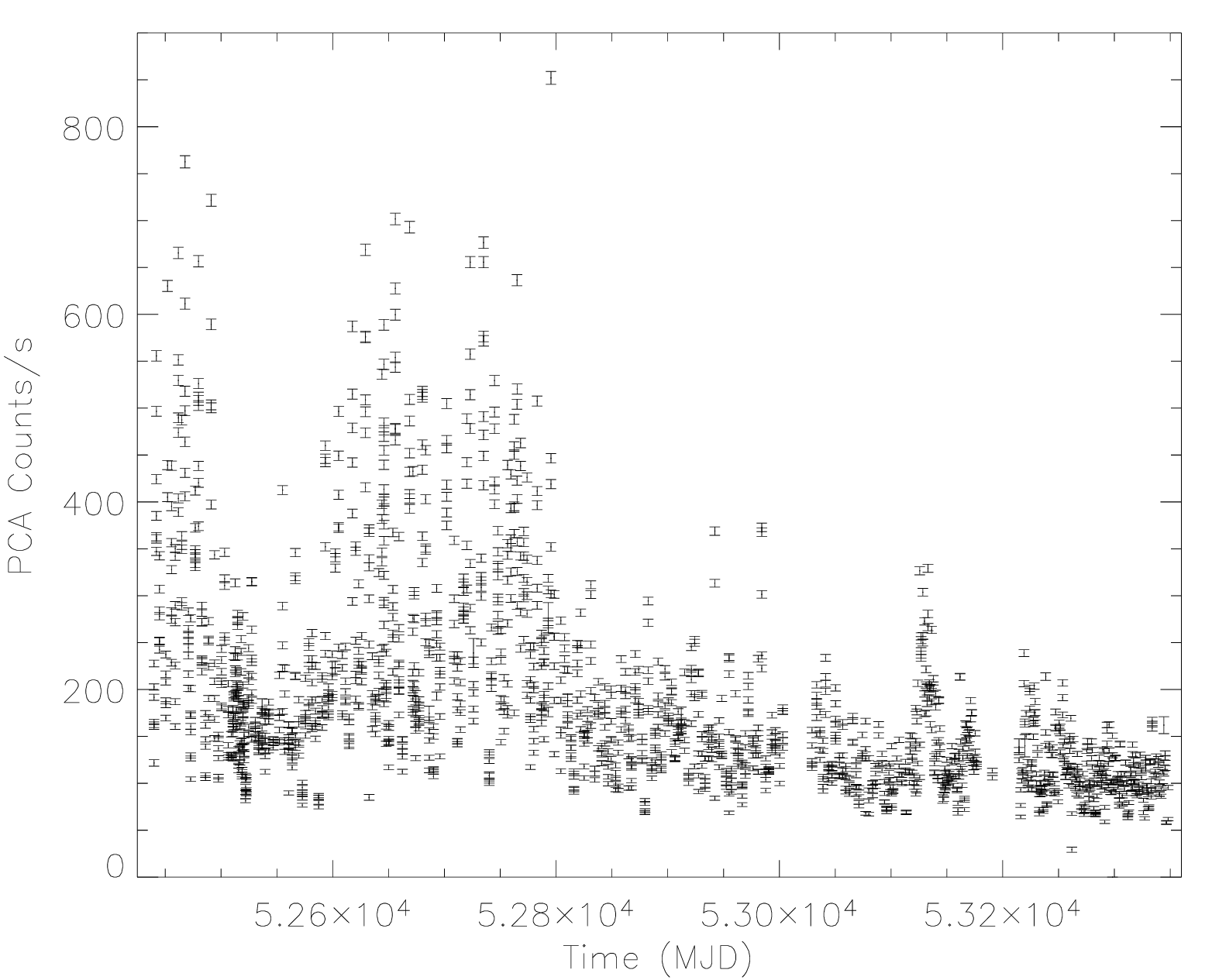,height=10cm,width=12cm}
\small{Fig. 1 -- {\it{RXTE-PCA}} light curve of SAX J2103.5+4545 after June 2002 outburst of the source.}
\end{center}
\end{figure}

Results of timing and spectral analysis of all the data after the 1999 outburst were presented before (Baykal et al. 2002; Baykal et al. 2000a,2000b). Inam et al. (2004a) analyzed $\sim 2$ months part of the data following the 2002 outburst to calculate the spin period and spin-up rate corresponding to the simultaneous {\it{RXTE}} - {\it{XMM-Newton}} observations of the source. 

In this paper, we extended our timing analysis with the observations after the 2002 outburst. The data after the 2002 outburst have a total exposure of $\sim 650$ ksec, and span the interval between July 2002 and December 2004 (see Figure 1). We also 
analyzed the evolution of the X-ray spectrum of the source in this interval.

\section{Timing Analysis}

\begin{figure}[bt]
\begin{center}
\psfig{file=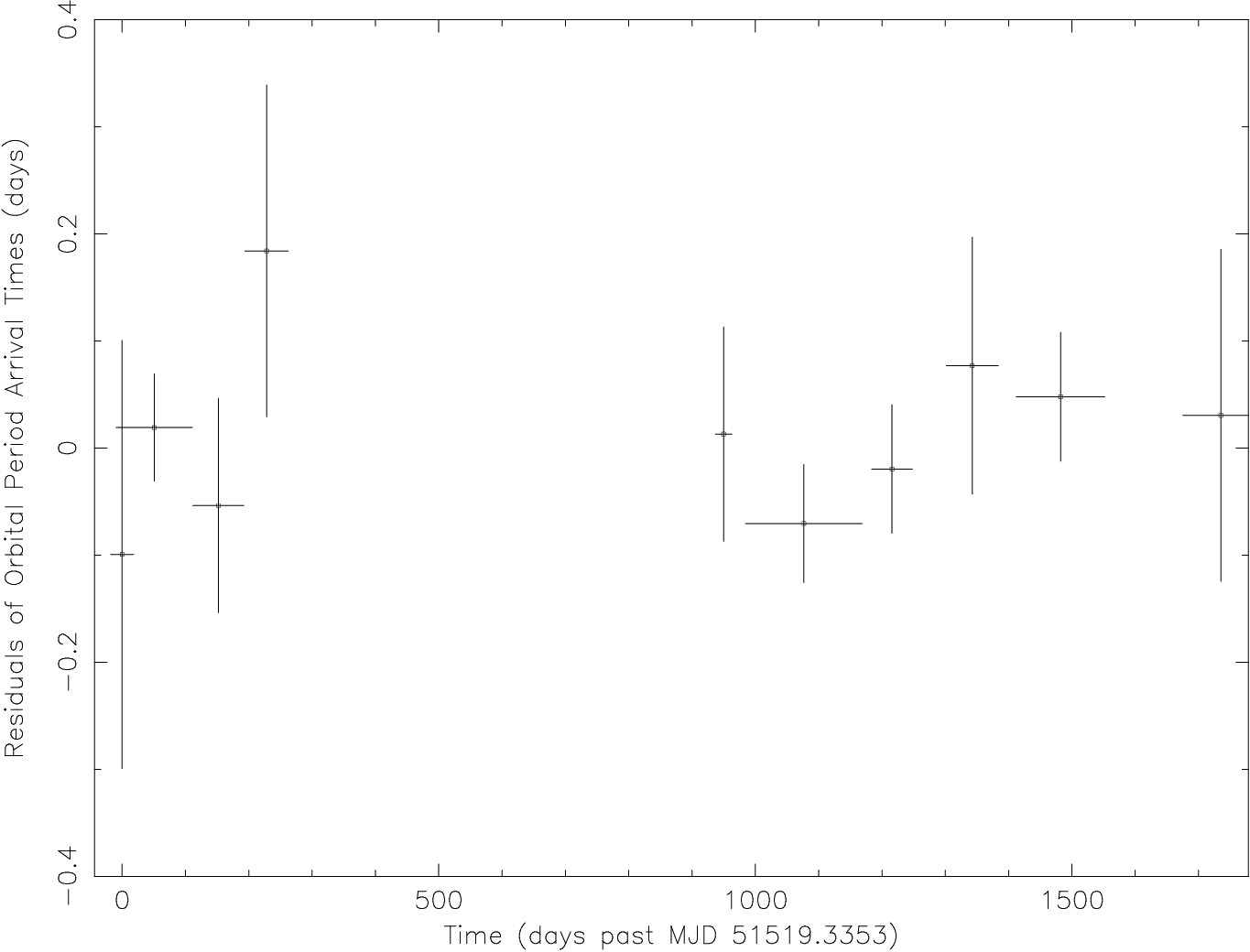,height=10cm,width=12cm,angle=-90}
\small{Fig. 2 -- Time delays of arrival times of orbital epochs
 of SAX J2103.5+4545 
}
\end{center}
\end{figure}

\begin{table}[bt]
\caption{Orbital epochs by pulse timing analysis}
\begin{center}
\begin{tabular}{c c c c } \hline
 Orbit Number & Orbital Epoch (MJD) 
 & Time span (days)& Reference \\ \hline
 0  & 51519.33$\pm$ 0.20  & 36  & Baykal et al., 2000  \\
 4  & 51570.11$\pm$ 0.05 & 120 & this work  \\
 12  &51671.36$\pm$ 0.10  & 80  &  this work\\
 18  &51747.59$\pm$ 0.15 & 68  &  this work\\
 75  &52469.35  $\pm$ 0.10  & 26  &  this work\\
 85  &52595.92  $\pm$ 0.05 & 184 &  this work\\
 96  &52735.29  $\pm$ 0.05 & 64  & this work\\
 106 &52862.04  $\pm$ 0.11 &  82 &  this work\\
 117 &53001.33  $\pm$ 0.05 & 140 & this work\\ 
 137 &53254.62  $\pm$ 0.15 & 120 &  this work \\ \hline
\end{tabular}
\end{center}
\end{table}

\begin{table}[tb]
\caption{Orbital Parameters of SAX J2103.5+4545}
\begin{center}
\begin{tabular}{c | c } \hline
Parameter & Value    \\ \hline
T$_{\pi/2}$ Orbital Epoch (MJD) & 52469.336 $\pm$0.057 \\
a$_{x}$ sin i (lt-s) & 74.07$\pm$0.86 \\
e  & 0.4055$\pm$ 0.0032 \\
$\omega$(deg) & 244.3$\pm$ 6.0 \\
$P_{orbit}$ & 12.66536$\pm$ 0.00088 \\
\end{tabular}
\end{center}
\end{table}

\begin{figure}[tb]
\begin{center}
\psfig{file=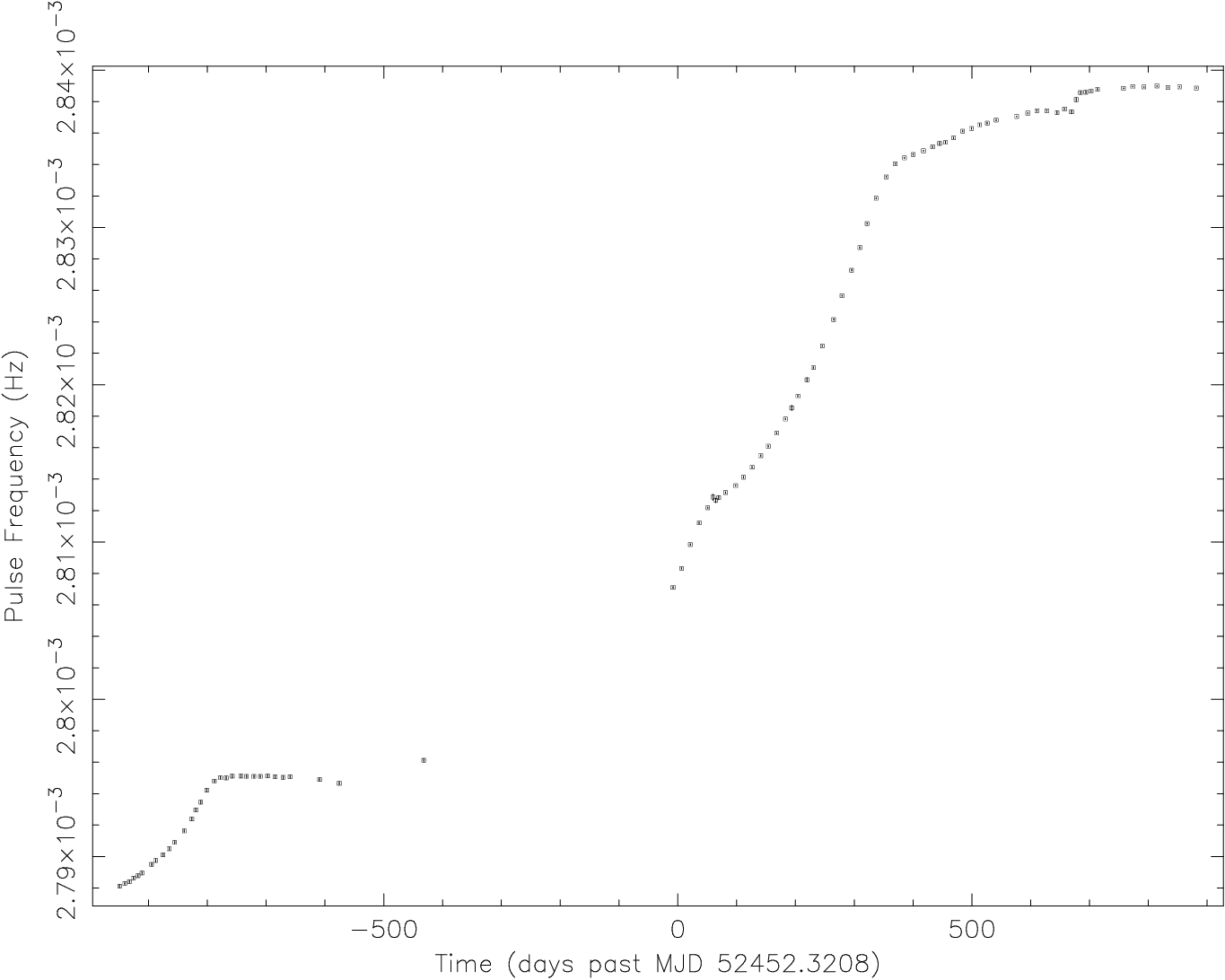,height=10cm,width=12cm,angle=-90}
\small{Fig. 3 -- Pulse Frequency History of  
  SAX J2103.5+4545
}
\end{center}
\end{figure}

\begin{figure}[bt]
\begin{center}
\psfig{file=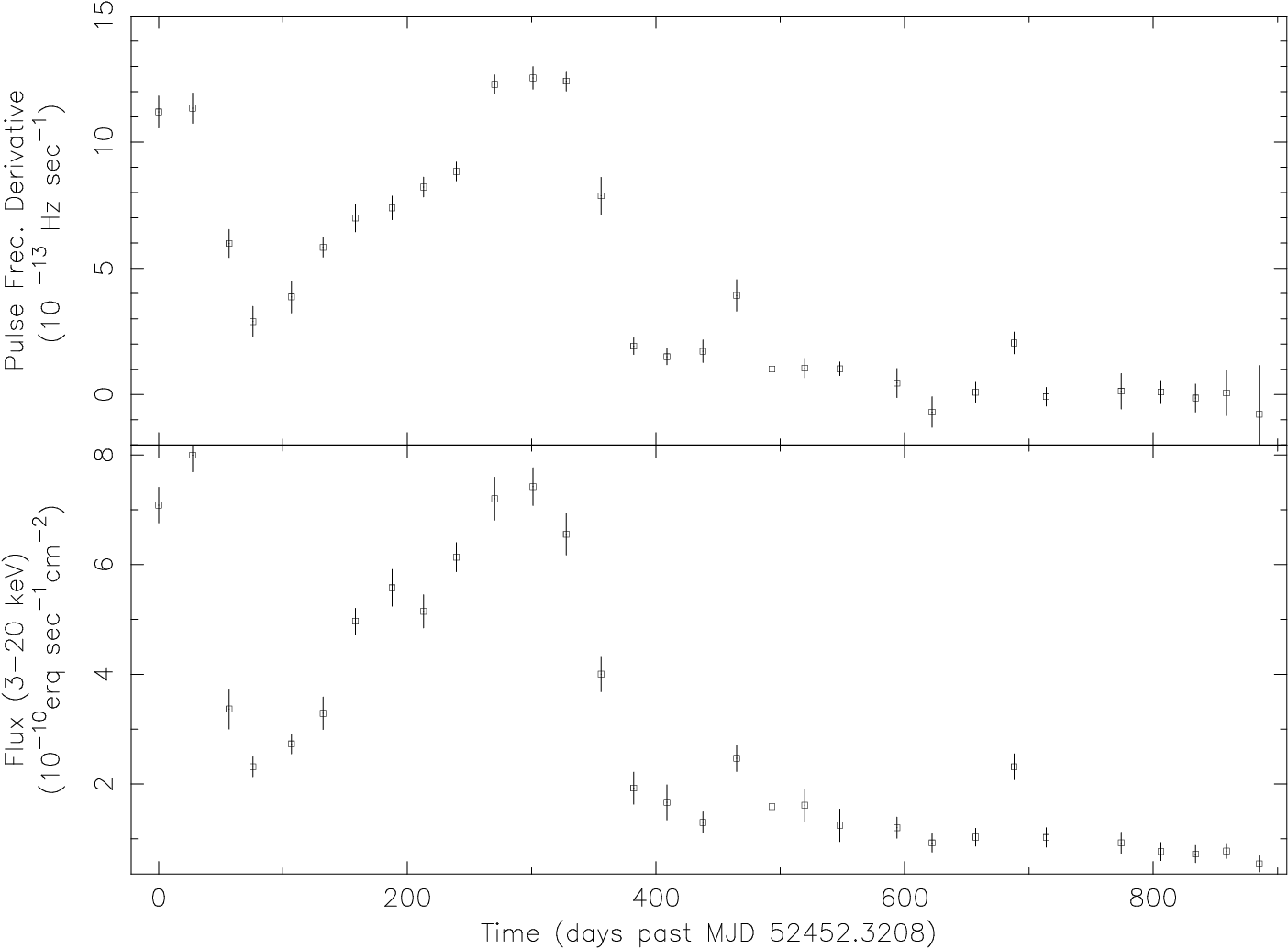,height=10cm,width=12cm,angle=-90}
\small{Fig. 4 -- Pulse Frequency Derivatives and 
X-Ray Flux history of 
  SAX J2103.5+4545
}
\end{center}
\end{figure}

\begin{figure}[bt]
\begin{center}
\psfig{file=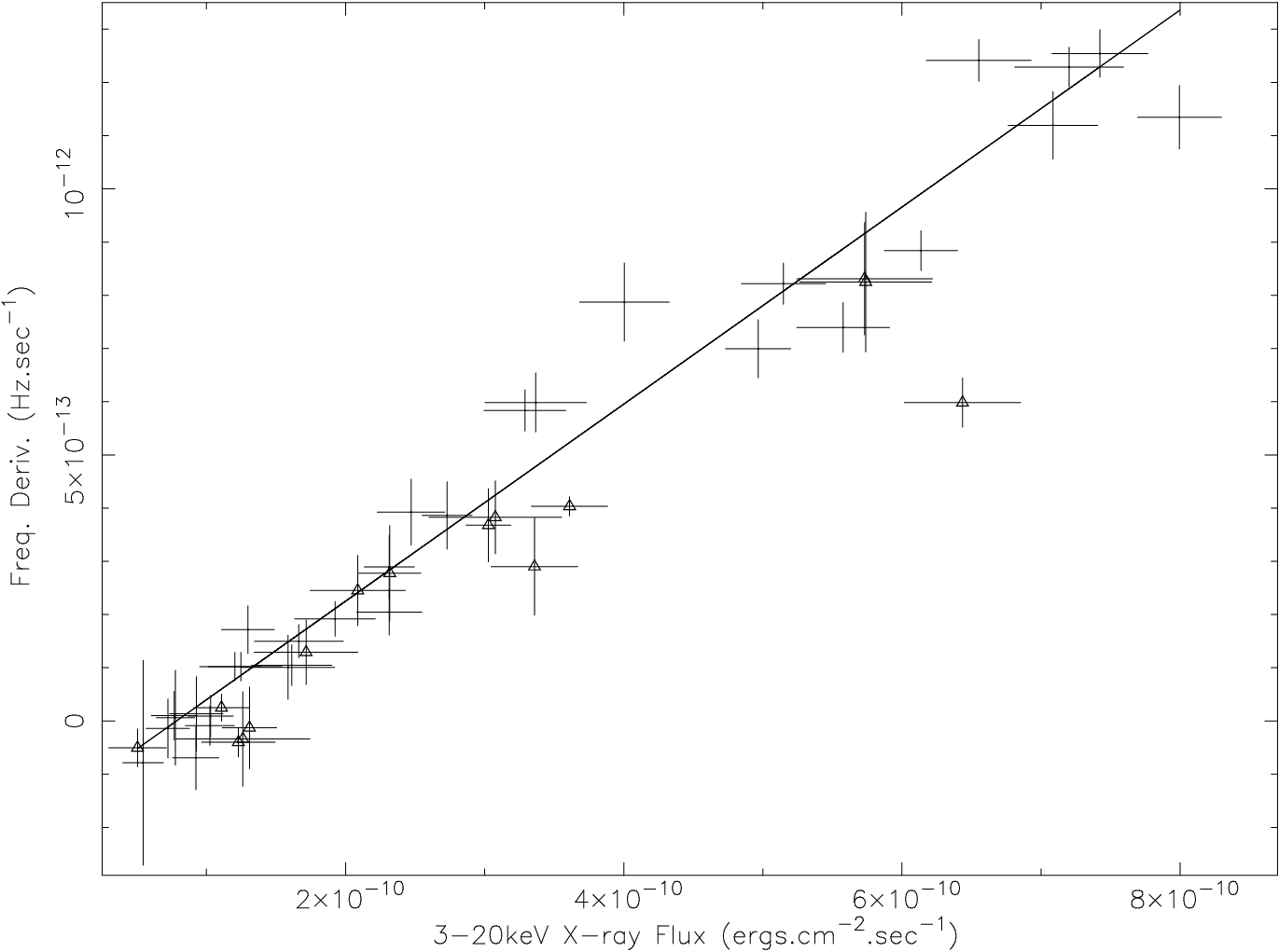,height=10cm,width=12cm,angle=-90}
\small{Fig. 5 -- X-ray flux and pulse frequency correlations, 
the best fit denotes Ghosh-Lamb model. Points denoted by triangles correspond 
to previous outburst's values found by Baykal et al. 2002 }
\end{center}
\end{figure}

\begin{figure}[hbt]
\begin{center}
\psfig{file=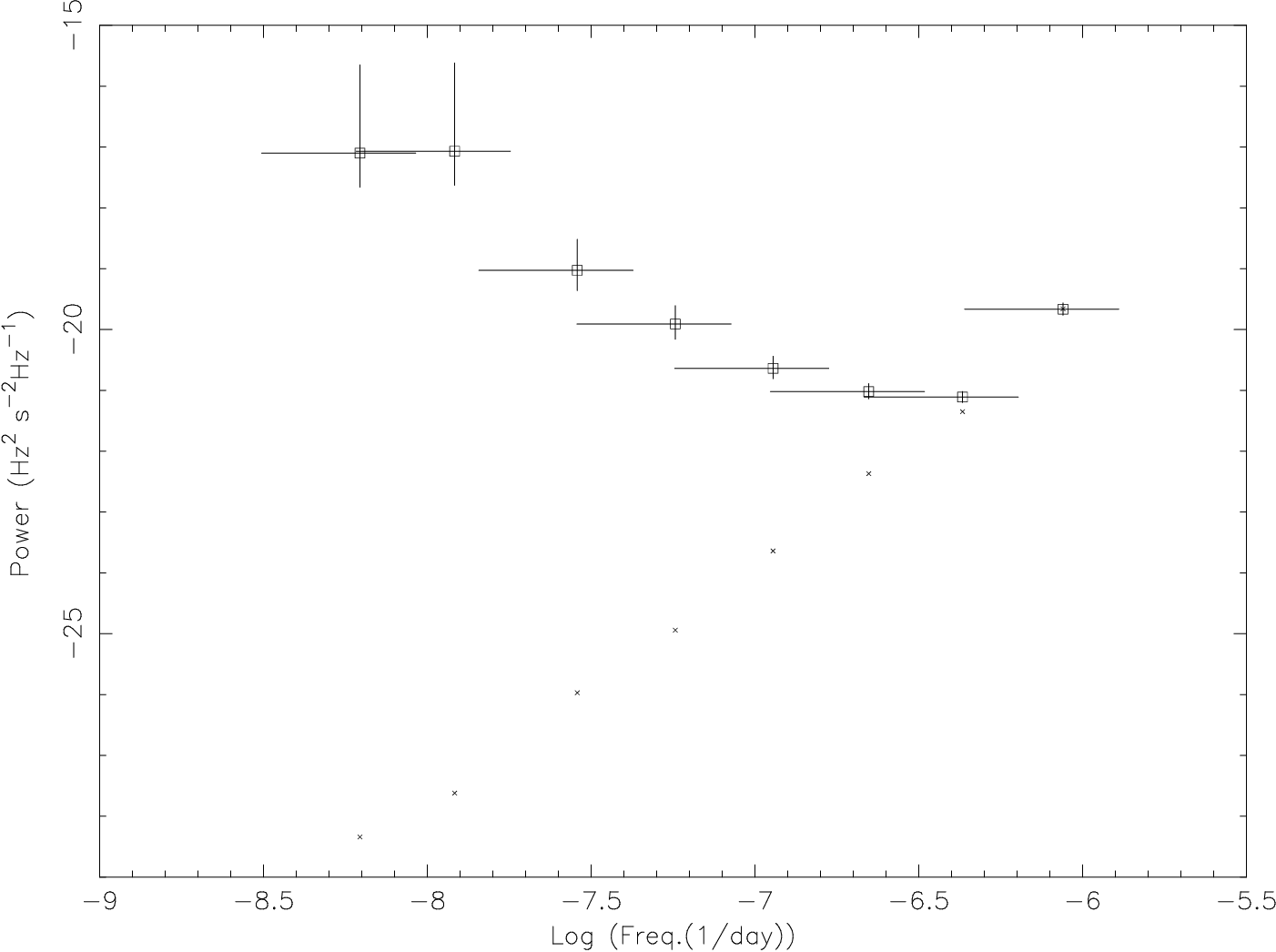,height=10cm,width=12cm,angle=-90}
\small{Fig. 6 -- Power Density of
 Pulse Frequency Derivatives of
  SAX J2103.5+4545, asterisk denotes 
the instrumental noise.
}
\end{center}
\end{figure}

Preliminary timing analysis was performed by folding the light curve on trial periods (Leahy et al., 1983) 
and producing a
chi-square test for each one day observation. The period corresponding to the maximum chi-square value gives the approximate value of the pulse period of 
SAX J2103.5+4545.  
We generated pulse profiles from each RXTE observation. 
Using the orbital parameters, and correcting for light travel time 
delays, we also constructed a template pulse profile.
We found pulse arrival times by 
cross-correlating the pulse profiles with the template pulse. 
In the pulse timing analysis, we used harmonic 
representation of pulse profiles, which was proposed by 
Deeter $\&$ Boynton (1985). 
In this method, the pulse profiles
for each orbit and the master profile are expressed in
 terms of harmonic series. We used 10 unweighted harmonic series to
cross-correlate the template pulse profile with the pulse profiles for each
RXTE observation. The maximum value
of the cross-correlation is analytically well-defined and
does not depend on the phase binning of the pulses.  

The source SAX J2103.5+4545 has a variable pulse profile which
affects the pulse timing. In order to estimate
 the errors in the arrival times,
the light curve of each RXTE observation was
 sampled into approximately 4-5 equal
subsets and new arrival times were estimated.
The standard deviation of the arrival time delays obtained
from each subset of the observation was taken to be the
uncertainty of the arrival time of that observation.

The observations were sampled a couple of times per week. Within these time intervals the contribution 
of pulse frequency derivatives on cycle counts 
 is not significant (i.e $1 << (1/2)  \dot \nu \Delta t)^{2}$) therefore 
we were able to phase connect the pulse arrival times for 
the data after November 1999 and June 2002  outbursts separately.
 Since there is a $\sim 450$ days gap between these two datasets, we did not phase-connect arrival times together. 
In order to refine the orbital period, we first
 obtained the orbital epochs 
($T_{\pi}/2$).
We divided pulse arrival times into 11 independent  
subsets. Time series corresponding to each  pulse arrival time 
had a time span between 36 and 184 days.
The  arrival time delays may arise from the
variations of the pulse frequency and its time derivatives (or
intrinsic pulse frequency derivatives) and variations of the
orbital parameters (Deeter, Boynton and Pravdo 1981, Boynton et l., 1986).

 In each interval, we fitted pulse phases $\phi $ 
(negative of the pulse arrival times normalized to pulse period)
to Taylor expansion and orbital model,
\begin{equation}
\delta \phi = \delta \phi_{o} + \delta \nu (t-t_{o})
+  \sum _{n=2}^{5} \frac{1}{n!}
 \frac {d^{n} \phi}{dt^{n}} (t-t_{o})^{n}
+ f(t_{n})
\end{equation}

where $\delta \phi $ is the pulse phase offset deduced from the pulse
timing analysis,  $t_{o}$ is the mid-time of the each data set;
$\delta \phi_{o}$ is the residual phase offset at t$_{o}$;
$ \delta \nu$ is the correction of pulse frequency at time $t_0$;
$ \frac {d^{n} \phi}{dt^{n}} $ for n=2,3,4,5
 are the first, second, third and fourth 
 order derivatives of pulse frequency;
f($t_{n}$) characterizes the orbital Doppler delay which is parametrized by 
five Keplerian orbital parameters; projected semi-major axis 
$a_{x}$/c sin i (where i is the inclination angle between the line of 
sight and the orbital angular momentum vector), orbital period $P_{orb}$, 
eccentricity e, longitude of periastron $\omega $ and orbital epoch 
$T_{\pi /2}$  which is defined when the mean longitude is 90.  
 We used fifth order polynomials in the Taylor expansion.
The degree of the polynomial is  completely arbitrary. We increased the degree of polynomial until we obtained the rms values of phase residuals at the
order of uncertainity of the phase estimates
($\sim 0.04$). We obtained the polynomial free trends in
phase residuals when we removed fifth order polynomials
trend from pulse phases for the time intervals given in Table 1.
 The time span of each set depends on pulse frequency fluctuations, for
less pulse frequency fluctuations we used longer time spans of pulse phase data sets. 

Then using the $\Delta \chi ^{2}$ method for single parameter estimation,  
we estimated the orbital epochs in 1$\sigma $ confidence level 
(Press et al., 1986, Bevington 1969). 
In Table 1, we present the orbital epoch measurements 
and orbital cycle number (n).
In Figure 2, we present observed minus calculated values
of orbital epochs
($T_{\pi/2}-n<P_{orbit}>-<T_{\pi/2}-n<P_{orbit}>>$) relative to the
constant orbital period ($<P_{orbit}>=12.66536$ days). As seen from
Figure 2, residuals of arrival times of orbital epoch is
consistent with a new value, i.e. 12.66536 days, of constant orbital period
A quadratic fit to the epochs yields
an estimate of upper limit for the rate of period change
$\dot P_{orb}/P_{orb} =\pm 1.4 \times 10^{-4}$ yr$^{-1}$.

Using the constant new orbital period, we fitted pulse phases
 to orbital parameters $a_{x}$/c sin i, eccentricity e and 
longitude of periastron $\omega $ 
simultaneously for all segments along with
an independent orbital epoch $T_{{\pi}/2}$ and pulse frequency derivatives
  $d^{n} \nu / dt^{n} $,
where n=2,3,4,5 for each segment.
 In Table 2, we present refined orbital parameters 
and their uncertainties.
 
Using this new refined orbital parameters, we regenerated
 the pulse phases and  
constructed the pulse frequency time series
 for all RXTE/PCA observations.
For the estimation of pulse frequencies, we made linear fits to the phase
 offsets  ($\delta \phi = \phi _{0} + \delta \nu (t-t_{0})$) with nearly 
one orbital period resolution. These are presented in Figure 3. 
We constructed the pulse frequency derivatives by adding a quadratic term
$(1/2) \dot \nu (t-t_{0})^{2}$ to the Taylor expansion. We estimated the 
pulse frequency derivatives from the coefficients of quadratic polynomials.
For this purpose we fitted quadratic polynomial
to phase offsets at two orbital period time resolution 
shown at the top panel of Figure 4. From top panel of Figure 4 (between $\sim 178$ and $\sim 309$ days), our pulse frequency derivative values corresponding to the $\sim 131$ day interval between MJD 52629.9 and MJD 52761.3 are consistent with the previous pulse period derivative measurement using INTEGRAL observations (Sidoli et al. 2005).

The bottom panel of Figure 4 shows the X-ray fluxes associated with the 
pulse frequency derivatives. These X-ray fluxes were obtained from the spectral analysis of the 3-20 keV PCA data corresponding to the same time intervals that were used to calculate the frequency derivatives of the source (see next section for the details of our spectral analysis).   
In Figure 5, we present the 
pulse frequency derivatives and X-ray fluxes together with the
published values by Baykal, Stark and Swank (2002).
It is seen that, for both outbursts, pulse frequency
derivatives are correlated with X-ray flux values. 

In order to see the statistical trend of pulse frequency derivatives, we 
constructed the power spectrum of the pulse frequency derivatives. 
We used the Deeter polynomial estimator method (Deeter 1984) to derive the power spectrum from the pulse frequency measurements.
This technique uses the polynomial estimators
instead of sinusoidal estimates for each time scale T.
The power density estimator $P_{\dot \nu}({f})$ is defined as 
$\int _{0} ^{\infty} P_{\dot \nu}({f}) = 
< (\dot \nu -  < \dot \nu >)^{2}>   $ where $< \dot \nu >$ means 
pulse frequency derivative of given analysis frequency.
In order to estimate the  power  density, we first 
divided the spin frequency measurements 
into time spans of duration T and fitted a quadratic polynomial in time. 
The observed time series was simulated by a Monte Carlo 
technique for a unit 
white noise strength defined as $P_{\dot \nu}({f})=1$ and  
fitted with a quadratic polynomial in time.  
Then the square of the second order term was normalized to 
the value obtained from Monte Carlo simulation (Deeter 1984, Cordes 1980). 
The logarithmic average of these estimators over the same time intervals
is the power density estimate. 
This procedure was repeated for different durations T 
to obtain a power spectrum.
The power due to instrumental noise was subtracted from the estimates 
and shown independently by cross symbols. The frequency response of 
each power density and instrumental noise estimates are presented at 
$f \sim 1/T$. In Figure 6, we present
power of pulse frequency derivatives ($P_{\dot \nu}({f})$)
per Hertz as a function of analysis frequency $f$.
The slope of the power spectrum between 1/1858 and 1/52 $d^{-1}$ 
yields a power law index
2.13 $\pm$ 0.6. This is the steepest power law index seen among the HMXRBs 
(Bildsten et al. 1997).
It should also be noted that SAX J2103.5+4545 is the first transient HMXRB 
for which the noise power spectrum is constructed. The power spectrum 
indicates that at short time scales  pulse frequency derivative fluctuations are less noisy. On contrary, at long time scales, pulse frequency derivative noise strengths are 
stronger. This could be qualitatively explained by well defined accretion disk 
at shorter time scales possessing low timing noise therefore
 power density spectrum becomes 
more steeper relative the other persistent HMXRBs.        

\begin{figure}[p]
\begin{center}
\psfig{file=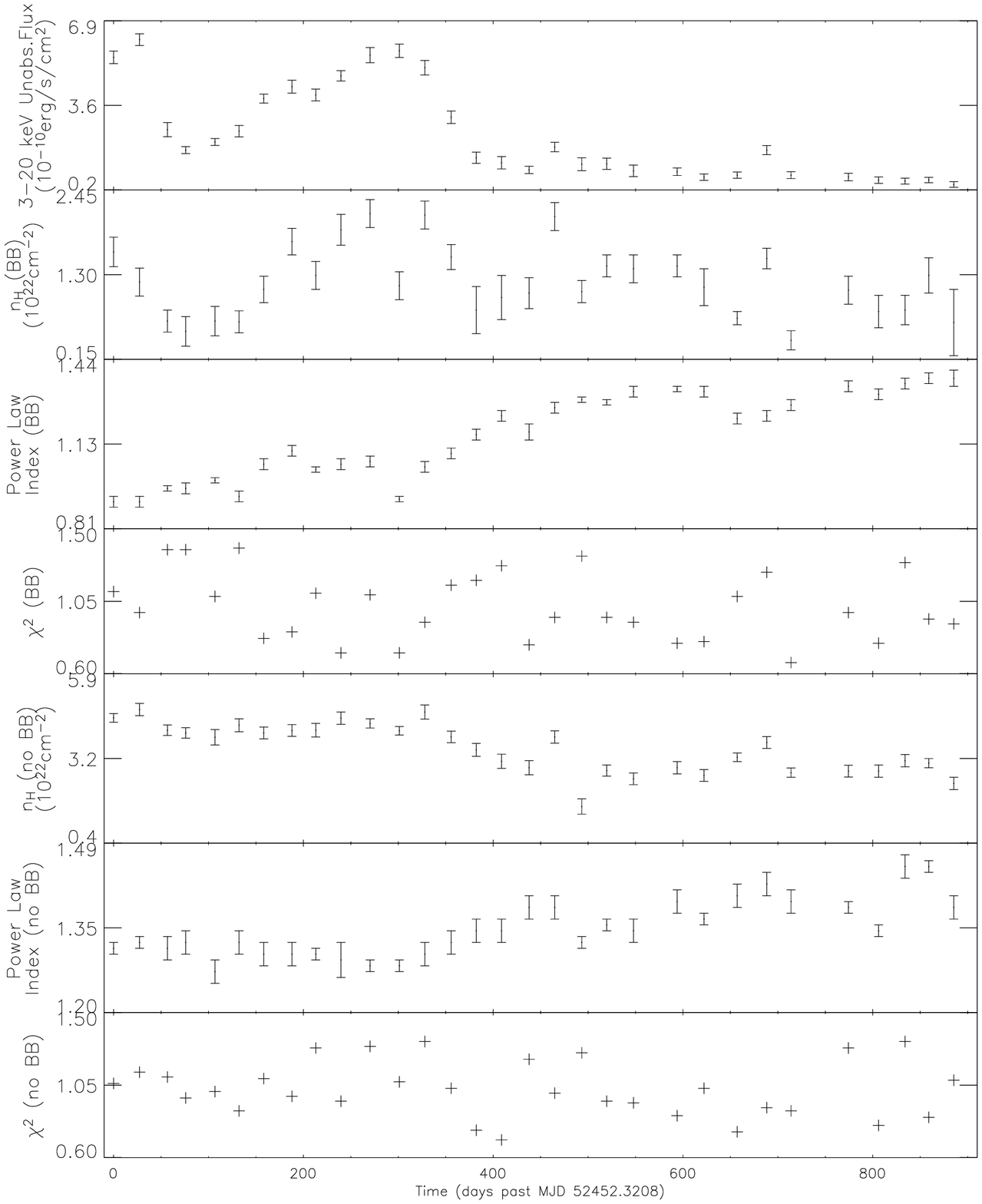,height=17cm,width=14cm}
\small{Fig. 7 -- Variation of 3-20 keV unabsorbed flux, Hydrogen column density, power law index and reduced $\chi^2$ for the spectral model with blackbody component, same spectral parameters and reduced $\chi^2$ for the spectral model without blackbody component. For the model with the blackbody component blackbody peak energy and iron line width were fixed at 1.9 keV and 0 keV respectively.}
\end{center}
\end{figure}

\section{Spectral Analysis}

We analyzed 31 PCA spectra of SAX J2103.5+4545 obtained from data corresponding to the same time intervals that were used to calculate the frequency derivatives shown in Figure 4.  

Spectrum, background and response matrix files were created using FTOOLS 6.0 data analysis software. We used background subtracted spectra in our analysis. Energy channels corresponding to the 3-20 keV energy range were used to fit the spectra. We ignored photon energies lower than 3 keV due to uncertainties in background modelling while energies higher than 20 keV were ignored as a result of poor counting statistics. No systematic error was added to the errors.

We fitted the spectra using two models. The first model consists of an absorbed power law (Morrison \& McCammon 1983) with a high-energy cutoff (White et al. 1983). The second model has an additional soft component modeled as a blackbody using BBODYRAD model in XSPEC software with the peak energy freezed at 1.9 keV. This frozen peak energy is consistent with the previously found value of the soft blackbody component of SAX J2103.5+4545 (Inam et al. 2004a). For both models, an iron line feature at $\sim 6.4$ keV was required to fit the spectrum. The first model was previously found to well-fit the X-ray spectrum of SAX J2103.5+4545 for the energies higher than $\sim 3$ keV using RXTE (Baykal et al. 2002; Inam et al. 2004a), and INTEGRAL (Lutovinov, Molkov,\& Revnivtsev 2003; Flippova, Lutovinov, Shtykovsky et al. 2004) observations. Using XMM-Newton observations, the second model was found to fit the spectrum better when the energies lower than $\sim 3$ keV were included (Inam et al. 2004a). 

Our results showed that two spectral models fit almost equally well to the spectrum in 3-20 keV band (see Figure 7). For the model with the blackbody component, cut-off and e-fold energy parameters were found to vary between $\sim 13.2-17.0$ keV and $\sim 13.3-34.0$ keV. For the model without the blackbody component, the same parameters were found to vary between $\sim 6.6-10.9$ keV and $\sim 17.0-45.1$ keV. 

For the blackbody model, the blackbody normalization was found to vary between $\sim 0.05-1.07$ km$^2$ (10kpc)$^{-2}$. Corresponding blackbody and power law fluxes were found to be correlated with each other. Thus, the total unabsorbed flux was found to be correlated with the blackbody flux.

\section{Discussion}

 A correlation between pulse frequency derivative and X-ray flux has been 
observed in the outbursts of many X-ray pulsar transient systems. 
These systems are EXO 2030+375 (Parmar, White $\&$ Stella 1989, Parmar et al. 
1989, Reynolds et al. 1996), 2S 1417-62 (Finger, Wilson  $\&$ Chakrabarty 1996, 
Inam et al., 2004b), 
GRO J1744-28 (Bildsten et al. 1997), XTE J1543-568 (In't Zand, Corbet 
$\&$ Marshall 2001), KS 1947+300 (Morgan et al. 2002; Tsygankov \& Lutovinov 2005) and SAX J2103.5+4545 
(Baykal, Stark  $\&$ Swank 2002). All of these sources have shown correlations 
between spin-up rate and X-ray flux. Among them only SAX J2103.5+4545 
has been observed to have both spin-up and spin-down episodes. The correlation 
between spin-up/down and X-ray flux can be explained by accretion from an
accretion disk. Both outbursts  of SAX J2103.5+4545 
started with a spin-up trend, made a transition to a steady spin rate 
and then appeared to just begin a spin-down trend (see Figure 4 and Baykal et al. 2002). 
In Figure 5, we present the X-ray flux and pulse frequency derivative 
correlations including the results found in this paper and in Baykal et al. (2002).

If the accretion is via an accretion disk then 
the Keplerian rotation of the disk is disrupted by the 
the magnetosphere at the inner 
disk edge. Then the plasma is forced to  
accrete along the magnetic field lines.

 The inner 
disk edge $r_{0}$ moves inward with increasing mass accretion rate. 
The dependence of the inner disk edge
$r_{o}$ on the mass accretion rate $\dot M$
may approximately be expressed as
(Pringle $\&$ Rees 1972, Lamb, Pethick, $\&$ Pines 1973)
\begin{equation}
r_{o}= K \mu^{4/7}(GM)^{-1/7}\dot M^{-2/7}
\end{equation}
where 
$\mu\simeq BR^{3}$ is the neutron star magnetic moment with
$B$ the magnetic field and  $R$ the neutron star radius,
G is the gravitational constant, and $M$ is the
mass of the neutron star. In this equation $K=0.91$
 gives the Alfven
radius for spherical accretion.
 Then the torque estimate is given by
 Ghosh \& Lamb (1979) as
 \begin{equation}
2\pi I \dot \nu = n(w_{s}) \dot M~l_{K},
\end{equation} 
where $I$ is the moment of inertia of the neutron star,
$l_{K} = (GM)r_{o})^{1/2}$ is
 the specific angular momentum added by a Keplerian disk
 to the neutron star at the inner disk edge
 $r_{o} $;
\begin{equation}
n(w_{s}) \approx 1.4 (1-w_{s}/w_{c})/(1-w_{s})
\end{equation}
 is a dimensionless
torque which is a measure of the variation of the accretion torque
as estimated by the fastness parameter
\begin{equation}
w_{s} =\nu /\nu _{K}(r_{o}) = (r_{o}/r_{co})^{3/2} =
 2 \pi 
K^{3/2} P^{-1}  (GM)^{-5/7}
    \mu ^{6/7} \dot M^{-3/7},
\end{equation}
where $r_{co}=(GM/(2\pi\nu)^{2})^{1/3}$ is the corotation radius
at which the centrifugal forces balances the gravitational forces,
$w_{c}$
is the critical fastness parameter at which the accretion
torque is expected to vanish.
The critical fastness parameter $w_{c}$ has been estimated to be  
$\sim$ 0.35
and depends on the electrodynamics of the disk 
(Ghosh \& Lamb 1979, Wang 1987, Ghosh 1993,
Torkelsson 1998, Li $\&$ Wickramasinghe 1998, Dai $\&$ Li 2006).

The accreted material
will produce X-ray emission at the neutron star surface with a luminosity expressed as
\begin{equation}
L =  \eta GM \dot M /R
\end{equation}
where $\eta \le 1 $ is the efficiency factor. 
From
Equations 2,3 and 6, the rate of spin-up is related to the X-ray
intensity through
\begin{equation}
\dot \nu \propto n(w_{s}) L^{6/7} =  n(w_{s}) (4 \pi d^{2} F)^{6/7},
\end{equation}
where $d$ is the distance to the source and $F$ is the X-ray flux.

The distance of the source from optical observations was found to be 
$6.5\pm 0.9$ kpc (Reig et al. 2004; Reig et al. 2005). Although cyclotron emission has not been seen 
from this source, the hard spectrum and cut off energy suggests
 that the  magnetic field is greater than $10^{12}$ Gauss 
(Lutovinov, Molkov,\& Revnivtsev 2003; Flippova, Lutovinov, Shtykovsky et al.
2004; Falanga, di Salva, Burderi et al. 2005; Sidoli, Mereghetti, Larsson et al. 2005).
We fitted the Ghosh \& Lamb model setting power law index, 
magnetic field and distance as a free parameter.
We obtained for the distance to the source  $4.5\pm 0.5$ kpc and for
the magnetic field $(16.5\pm 2.5)\times 10^{12}$ Gauss.
We obtained a power-law index
$0.96 \pm 0.14$ which is consistent with the value 6/7 expected in the
model. The large uncertainty in this power law index shows that 
radiation pressure dominated accretion disk model with 
power law index 0.925 is also possible (see Ghosh 1996).
Since the peak outburst luminosity ($\sim 2\times 10^{36}$erg/sec, assuming a distance of 4.5kpc) is far 
below the Eddington limit ($\sim 10^{38}$ erg/sec), we consider that 
gas pressure accretion disk with power law index 6/7 
model is  the appropriate model for this source. 
It is also seen in Figure 5, the points corresponding to spin-up and spin-down
are consistent with torque model.

Our distance estimate is lower by $\sim 2 \sigma$  than the value obtained from optical observations. The discrepancy could arise from the efficiency factor
between accretion X-ray luminosity and observed luminosity. 
Another factor could be the uncertainty of 
 interstellar absorption coefficient used in optical  
observations (see Reig et al. 2004 and references therein)

The noise torque fluctuations of  accreting X-ray pulsars are subject to both 
external and internal torques. The external torques arise from both the
 accretion of matter from the binary companion and the interaction of the 
pulsar magnetosphere with accretion disk. The internal torques arise 
from the response of the neutron star to these external 
torques and depend on the interior structure of neutron stars. 
However due to the pulse shape noise and consequent instrumental 
noise, torque fluctuations at short time scales (or analysis frequencies) 
could not be resolved so far (Baykal, Alpar, Kiziloglu 1991; Baykal 1997). 
 Bildsten et al., (1997) constructed noise power spectra of HMXRBs using 
BATSE/CGRO pulsar observations. They found that for Vela X-1, Her X-1, 
4U 1538-52 and GX 301-2 pulse frequency derivative fluctuations   
($P_{\dot \nu}$) are consistent with a white torque noise model.
For OAO 1657-415, GX 1+4 and Cen X-3, they showed that red torque noise for which $P_{\dot \nu}$  is proportional to $f^{-1}$. 
Our results indicate that for SAX J 2103.5+4545 $P_{\dot \nu}$ is proportional to  $f^{-2}$. 
In other words torque fluctuations are characterized as first order random 
walk in pulse frequency derivatives. The main reason of this might be the 
 formation of  
transient accretion disks which leads to the formation of step like torque 
fluctuations on SAX J2103.5+4545. For higher analysis frequencies of 
$\sim 1/13 d^{-1}$, instrumental noise due to the pulse shape fluctuations 
limits the pulse timing and hence cut-off the torque noise fluctuations.

It was shown for SAX J2103.5+4545 that there is
 $\sim 50$\% X-ray flux
modulation within the time span of orbital period at the bright part
of observations
(see Figure 8 of Baykal, Stark, Swank 2000,
and Figure 4 of Baykal, Stark, Swank 2002).
Assuming that the pulse frequency and X-ray spectra correlations 
as seen in Figure 5 is
valid at short time scales in single orbit  
 we estimate the
phase offsets using the Ghosh and Lamb torque model 
as in  $\delta \phi \sim \int \int \dot \nu dt$
$\sim \int \int  L^{6/7} dt \sim
  0.0027 $. This phase offset corresponds to 
 the 1/0.0027=358 part of pulse phase. 
Typical uncertainty of pulse phases is $\sim 0.04$ which is 
much larger than the phase offsets of torque fluctuations.
Therefore we could not resolve the torque fluctuations 
within the orbital period time scale.

From the X-ray spectral analysis of SAX J2103.5+4545, we found that two spectral models fit almost equally well to data. The first spectral model consisting of absorbed power law with an high energy cut-off is a typical model for accretion powered pulsars (White et al. 1983). This spectral model has been found to well fit X-ray spectrum of SAX J2103.5+4545 before (Hulleman et al. 1998; Baykal et al. 2002; Lutovinov et al. 2003; Flippova et al.
2004; Play et al. 2004; Falanga et al. 2005; Sidoli et al. 2005)  

The second spectral model has an additional soft component modeled as a blackbody. The model with a soft component was recently found to better fit the X-ray spectrum of SAX J2103.5+4545  when energies lower than 3 keV 
were included (Inam et al. 2004a). While fitting the spectra using this model, we fixed the blackbody peak energy at 1.9 keV (typical value found by Inam et al. 2004a) since our energy range does not cover around 1.9 keV and large uncertainties of this parameter might have arised. The blackbody component in accretion powered pulsars comes either from the reprocessed emission of the surrounding material or from the polar caps of the neutron stars (see Inam et al. 2004a and references therein). Inam et al. 2004a found that the polar cap interpretation is more likely for the blackbody emission of SAX J2103.5+4545. For the distance of 4.5kpc and considering the variation of the blackbody normalization, we estimated the radius of blackbody emitting region to be varying between $\sim 0.01-0.47$ km which is consistent with $\sim 0.3$ km value found by Inam et al. 2004a.  

For both of the spectral models, we found that power law index anti-correlates with the unabsorbed X-ray flux  (see Figure 7). This anti-correlation indicates that the spectrum gets softer with the decreasing X-ray flux. This anti-correlation was previously found in the dataset after the 1999 outburst of SAX J2103.5+4545 (Baykal et al. 2002), and in the dataset after the 1999 outburst of 2S 1417-62 (Inam et al. 2004b). This type of spectral softening accompanied with decreasing flux was found to be primarily a consequence of mass accretion rate change and is not necessarily related to a significant accretion geometry change (Mezsaros et al. 1983; Harding et al. 1984). 

{\bf{Acknowledgments}}

S.\c{C}.\.{I} acknowledges research project TBAG 105T443 of the Scientific and Technological Research Council of Turkey (T\"{U}B\.{I}TAK).

\noindent{{\bf{References}}}

\noindent{Baykal, A., Alpar, A., Kiziloglu, U. 1991, A\& A, 252, 664}

\noindent{Baykal, A., 1997, A\& A, 319, 515}

\noindent{Baykal, A., Stark, M., Swank J. 2000a, IAU Circ. 7355}

\noindent{Baykal, A., Stark, M., Swank J. 2000b, ApJ, 544, L129}

\noindent{Baykal, A., Stark, M., Swank J. 2002, ApJ, 569, 903}

\noindent{Bevington, P.R. 1969, "Data Reduction and error analysis for the physical sciences", McGraw-Hill}

\noindent{Bildsten, L., Chakrabarty, D., Chiu, J. et al. 1997, ApJS, 113, 367}

\noindent{Blay, P., Reig, P., Martinez Nunez, S., Camero, A., Connell, P., Reglero, V. 2004, A\& A, 427, 293}

\noindent{Boynton, P.E., Deeter, J.E., Lamb, F.K., Zylstra, G. 1986, ApJ, 307, 545}

\noindent{Cordes, J.M. 1980, ApJ, 237, 216}

\noindent{Dai, H.-L., Li, X.-D. 2006, A\& A, 451, 581}

\noindent{Deeter, J.E. 1984, ApJ, 281, 482}

\noindent{Deeter,  J.E., Boynton,  P.E., 1985, in Proc. Inuyama Workshop 
on Timing Studies of X-Ray Sources, ed. S. Hayakawa $\&$ F. Nagase 
(Nagoya: Nagoya Univ.), 29}

\noindent{Deeter, J.E., Pravdo, S.H., Boynton, P.E., 1981, ApJ, 247,1003}

\noindent{Falanga, M., di Salvo, T., Burderi, L. et al. 2005, A\& A, 436, 313} 

\noindent{Filipova, E.V., Lutovinov, A.A., Shtykovsky, P.E., Revnivtsev, M.G., Burenin, R.A., Arefiev, V.A., Pavlinsky, M.N., Sunyaey, R.A. 2004, AstL, 30, 824}

\noindent{Finger, M.H., Wilson, R.B., Chakrabarty, D. 1996, A\& AS, 120, 209}

\noindent{Ghosh, P., Lamb F.K. 1979, ApJ, 234, 296}

\noindent{Ghosh, P. 1993, in Holt S.S., Day, C.S. eds. The Evolution of X-ray Binaries, Am.Inst. Phys., New York, p.439}

\noindent{Harding, A.K., Kirk, J.G., Galloway, D.J., Meszaros, P. 1984, ApJ, 278, 369}

\noindent{Hullemann, F., in't Zand, J.J.M., Heise, J. 1998, ApJ, 337, L25}

\noindent{Inam, S.C., Baykal, A., Swank, J., Stark, M.J. 2004a, ApJ, 616, 463}

\noindent{Inam, S.C., Baykal, A., Scott, D.M., Finger, M., Swank, J. 2004b, MNRAS, 349, 173}

\noindent{in't Zand, J.J.M., Corbet, R.H.D., Marshall, F.E. 2001, ApJL, 553, 165} 

\noindent{Jahoda, K., Swank J., Giles A.B. et al. 1996, Proc. SPIE, 2808, 59}

\noindent{Lamb, F.K., Pethick, C.J., Pines, D. 1973, ApJ, 184, 271}

\noindent{Li, J., Wickramasinghe, D.T. 1998, MNRAS, 300, 1015}

\noindent{Lutovinov, A.A., Molkov, S.V., Revnivtsev, M.G. 2003, AstL, 29, 713}

\noindent{Meszaros, P., Harding, A.K., Kirk, J.G., Galloway, D.J. 1983, ApJ, 266, 33}

\noindent{Morgan, E., Galloway, D.K., Chakrabarti, D., Levine, A.M. 2002, AAS, 201, 5401} 

\noindent{Morrison, R., McCammon, D. 1983, ApJ, 270,119}

\noindent{Parmar, A.N., White, N.E., Stella, L., Izzo, C., Ferri, P. 1989, ApJ, 338, 359}

\noindent{Parmar, A.N., White, N.E., Stella, L. 1989, ApJ, 338, 373}

\noindent{Press, W.H., Flannery, B.P., Teukolsky, S.A. 1986, "Numerical Recipes. The Art of Scientific Computing", Cambridge University Press}

\noindent{Pringle, J.E., Rees, M.J. 1972, A\& A, 21, 1}

\noindent{Reig, P., Negueruela, I., Fabregat, J., Chato, R., Blay, P., Mavromatakis, F. 2004, A\& A, 421, 673}

\noindent{Reig, P., Negueruela, I., Papamastorakis, G., Manousakis, A., Kougenakis, T. 2005, A\& A, 440, 637} 

\noindent{Reynolds, A.P., Parmar, A.N., Stollberg, M.T., Verbunt, F., Roche, P., Wilson, R.B., Finger, M.H. 1996, A\& A, 312 872}

\noindent{Sidoli, L., Mereghetti, S., Larsson, S. et al. 2005, A\& A, 440, 1033}

\noindent{Torkelsson, U. 1998, MNRAS, 298, 55}

\noindent{Tsygankov, S.S., Lutovinov, A.A. 2005, AstL, 31, 88}

\noindent{Wang, Y.-M. 1987, A\& A, 183, 257}

\noindent{White, N.E, Swank, J.H., Holt, S.S. 1983, ApJ, 270,711}

\end{document}